\documentclass[notitlepage,nofootinbib,preprintnumbers,amssymb,superscriptaddress]{revtex4-1}
\usepackage{amsfonts,amssymb,amsmath,graphicx,color,bm}
\usepackage[linktocpage=true]{hyperref}
\hypersetup{
colorlinks=true,
citecolor=red,
linkcolor=blue,
urlcolor=blue,
}

\begin{document}

\title{Generalized scalar-tensor theory of gravity reconstruction from physical potentials of a scalar field}

\author{I.V. Fomin}
\email[E-mail:]{ingvor@inbox.ru}
\affiliation{Department of Physics, Bauman Moscow State Technical University,
  Moscow, 105005, Russia}

\author{S.V. Chervon}
\email[E-mail:]{chervon.sergey@gmail.com}
\affiliation{Department of Physics, Bauman Moscow State Technical University,
  Moscow, 105005, Russia}
 \affiliation{Laboratory of  gravitation, cosmology, astrophysics, Ulyanovsk State Pedagogical University, Lenin's Square 4/5, Ulyanovsk 432071, Russia}
\affiliation{Institute of Physics, Kazan Federal University, Kremlevskaya street 18, Kazan, 420008, Russia}

\author{A.V. Tsyganov}
\email[E-mail:]{andrew.tsyganov@gmail.com}
\affiliation{Laboratory of  mathematical modeling, \\ Ulyanovsk State Pedagogical University, Lenin's Square 4/5, Ulyanovsk 432071, Russia}

%\preprint{1234}

\begin{abstract}
We describe how to reconstruct generalized scalar-tensor gravity (GSTG) theory, which admits exact solutions for physical type of the potentials. Our consideration deals with cosmological inflationary models based on GSTG with non-minimal coupling of a (non-canonical) scalar field to the Ricci scalar.
The basis of proposed approach to the analysis of these models is a priori specified relation between the Hubble parameter $H$ and a function of non-minimal coupling $F=1+\delta F$ as $H\propto\sqrt{F}$. Deviations from Einstein gravity $\delta F$ induce a corresponding deviations of the potential $\delta V$ from a constant value and modify the dynamics from pure de Sitter exponential expansion. We analyze the models with exponential power-law evolution of the scale factor and we find the equations of influence of non-minimal coupling, choosing it in the special form, on the potential and kinetic energies. Such consideration allows us to substitute the physical potential into obtained equations and then to calculate the non-minimal coupling function and kinetic term that are define GSTG parameters. With this method, we reconstruct GSTG for polynomial, exponential, Higgs, Higgs-Starobinsky and Coleman-Weinberg potentials.
Special attention we pay to parameters of cosmological perturbations and prove correspondence obtained solutions to observational data from Planck.
\end{abstract}

\maketitle

\section{Introduction}

Currently, a large number of cosmological inflationary models are being studied on the basis of various types of exotic matter~\cite{Frieman:2008sn, Unnikrishnan:2013vga,Chervon:2014dya} and modifications of Einstein's theory of gravity~\cite{Nojiri:2010wj,Nojiri:2017ncd,Clifton:2011jh,Elizalde:2004mq,Chervon:2019jfu}. These models of the early universe can be considered as a development of the first cosmological inflation models based on the evolution of a certain scalar field and Einstein's gravity~\cite{Starobinsky:1980te,Guth:1980zm,Linde:1981mu,Albrecht:1982wi,Chervon:2019sey}.
The reasons for modifying the original inflationary models are the need to build a quantum theory of gravity~\cite{Baumann:2014nda} and explain the repeated accelerated expansion of the universe in the present era~ \cite{Perlmutter:1998np,Riess:1998cb}.

One of the first modifications of GR is the scalar-tensor gravity theories~\cite{Fujii:2003pa,Faraoni:2004pi}. This type of modified gravity theories successfully solves the problems of constructing consistent scenarios for the evolution of the early universe and its repeated accelerated expansion. It should also be noted that the speed of propagation of gravitational waves in the case of scalar-tensor theories of gravitation is equal to the speed of light in vacuum, as in the case of Einstein's gravity~\cite{DeFelice:2011jm,DeFelice:2011bh}, which corresponds to recent measurements of  characteristics of gravitational waves from black holes and neutron stars  mergers~\cite{Monitor:2017mdv}.

There are various types of scalar-tensor theories of gravity, one of which is the case of non-minimal coupling of a material scalar field and curvature. The construction of inflationary scenarios based on these theories of gravity has been carried out in many studies using exact and approximate solutions of the equations of cosmological dynamics~\cite{Fujii:2003pa,Faraoni:2004pi,Belinchon:2016lwr,Pozdeeva:2016cja,Fomin:2017sbt,Fomin:2018blx,Fomin:2019yls}. The main criterion for the correctness of the constructed models of the early universe based on the inflationary paradigm is the compliance of the spectral parameters of cosmological disturbances with the observational constraints obtained from measurements of the anisotropy and polarization of CMB~\cite{Ade:2015xua}.

We note that the main parameters characterizing the type of cosmological model of the early universe are the potential of a scalar field $V(\phi)$, the non-minimal coupling function $F(\phi)$, which determine the physical processes that occur at the inflationary stage, and the Hubble parameter $H(t)$ corresponding to the dynamics of accelerated expansion. Therefore, it is possible to determine the relations between these parameters that gives the models corresponding to the observational data for some class of cosmological models.

In this paper, we study cosmological models based on the relation $H=\lambda\sqrt{F}$ with exponential power-law (EPL) dynamics $H(t)=\frac{m}{t}+\lambda$. This case is interesting in that it is a combination of de Sitter and Friedmann solutions, which implies a transition between these stages under certain conditions.

The determination of a priori given relations between the parameters of cosmological models or ``ansatzes" is often used to analyze them both for the case of Einstein gravity~\cite{Chervon:2019sey,Chervon:2017kgn,Martin:2012pe,Motohashi:2017aob} and its modifications~\cite{Motohashi:2017vdc,Motohashi:2019tyj,Fomin:2017qta,Fomin:2018typ,Chervon:2019nwq} as well. On the one hand, the use of ansatzes makes it possible to obtain exact analytical solutions of the dynamic equations, on the other hand, on the basis of this method it is possible to single out cases that have the correct physical interpretation.

As example of such an approach, we can mention the models of cosmological inflation with ``constant roll" based on GR, $f(R)$-gravity and scalar-tensor gravity ~\cite{Martin:2012pe,Motohashi:2017aob,Motohashi:2017vdc,Motohashi:2019tyj}. It is also possible to give many other examples of the use of anasatzes to construct and analyze the cosmological models (for example, see~\cite{Chervon:2019sey,Chervon:2017kgn,Fomin:2017qta,Fomin:2018typ,Chervon:2019nwq}).

The physical content of the relation $H=\lambda\sqrt{F}$ in the inflationary models based on GSTG is that the non-minimal coupling of a scalar field and curvature, otherwise the deviations from Einstein gravity, induce the corresponding deviations from pure exponential expansion and the deviations of it's potential from the constant~\cite{Fomin:2017sbt}. Also, for any potential of a scalar field in these models, the observational constraints on the spectral parameters of cosmological perturbations will be satisfied. Thus, the initial problem of analyzing inflationary models in this case is reduced to reconstructing the parameters of the scalar-tensor gravity theory and the evolution of the scalar field for a known potential which is the main objective of this article.

The article is organized as follow. In Sec.\ref{sect2} we represent the action of the GSTG model and its connection to GR with scalar field. The dynamic equations in Friedmann-Robertson-Walker metric for GSTG are displayed.
Sec.\ref{sect3} contains the choice of the quadratic connection between Hubble parameter and coupling function of the specific kind, which then applied for obtaining general form of solution for exponential-power-law expansion. Accelerated period, exit from early inflation and the second accelerated expansion stage are studied for such expansion.
The exact solutions for generalized polynomial potential and for it's partial cases (polynomial and exponential) are represented in Sec.\ref{sect4}. All cosmological parameters are defined by the choice of one generating function $\Psi(\phi)$ for these EPL inflationary models.
In Sec.\ref{sect6} we represent the reconstructed models from physically motivated potentials, namely Higgs, Higgs-Starobinsky, Coleman-Weinberg and massive scalar field potentials.
Sec.\ref{sect5} devoted to calculation of cosmological parameters for the model under consideration and contains the prove of satisfaction the model’s prediction for observational restrictions. As the result, we obtained that EPL-models based on GSTG under consideration correspond to observational constraints on the parameters of cosmological perturbation for any $m>0$  and estimated the value of the parameter $\lambda$ as well.
In Sec.\ref{sect7} we summing up the results of our investigation.

\section{The model and cosmology equations}\label{sect2}

The generalized scalar-tensor theory is described by the action~\cite{Fujii:2003pa,Faraoni:2004pi}
\begin{eqnarray}\label{actionFR}
S_{(STG)} = \int d^4x\sqrt{-g}\Big[\frac{1}{2}F(\phi) R -
 \frac{\omega(\phi)}{2}g^{\mu\nu}\partial_{\mu}\phi \partial_{\nu} \phi
 - V(\phi)\Big] +S^{(m)},\\
S^{(m)}=\int d^4x\sqrt{-g} {\cal L}^{(m)},
\end{eqnarray}
where Einstein gravitational constant  $ \kappa=1$, $g$ is the determinant of space-time metric tensor $g_{\mu\nu}$, $\phi$ is a scalar field with the potential $V=V(\phi)$, $\omega(\phi)$ and $F(\phi)$ are the differentiable functions of $\phi$ which define the type of gravity theory, $R$ is the Ricci scalar of the space-time, ${\cal L}^{(m)}$ the Lagrangian of the matter.
In the present article we will study the case of vacuum solutions for the model (\ref{actionFR}) suggesting that $ S^{(m)}=0$.

The action of the scalar field Einstein gravity is
\begin{equation}\label{actionR}
S_{(GR)} = \int d^4x\sqrt{-g}\left[\frac{R}{2} -
\frac{1}{2}g^{\mu\nu}\partial_{\mu}\phi \partial_{\nu} \phi-V(\phi)\right].
\end{equation}

Let us remind that the scalar field $\phi$ is a gravitational field in the action (\ref{actionFR}), while the scalar field $\phi$ in the action (\ref{actionR}) is the source of the gravitation and it is non-gravitational (material) field.

Considering the gravitational and scalar field equations of the models (\ref{actionFR}) and (\ref{actionR}) in the Friedman universe we use the choice of natural units, including $ \kappa=1$. Therefore we could not make difference between gravitational and non-gravitational (material) scalar fields. Thus, the solutions of the model's equations should be related to the subsequent situation. That is, from the physical point of view, the solutions will correspond to different theories of gravity.

Let as note also that the cosmological constant $\Lambda$ can be extracted from the constant part of the potential $V(\phi)$, therefore we did not include it into the actions (\ref{actionFR}) and (\ref{actionR}).

To describe a homogeneous and isotropic universe we chose the Friedmann-Robertson-Walker (FRW) metric in the form
\begin{equation}\label{FRW}
ds^2=-dt^2+a^2(t)\left(\frac{d r^2}{1-k r^2}+r^2 \left( d\theta^2+\sin^2\theta d\varphi^2\right)\right),
\end{equation}
where $a(t)$ is a scale factor, a constant  $k$ is the indicator of universe's
type:
$ k>0,~k=0,~k<0 $ are associated with closed, spatially-flat, open universes,  correspondingly.

The cosmological dynamic equations for the STG theory (\ref{actionFR}) in a spatially-flat FRW metric are~\cite{DeFelice:2011jm}
\begin{eqnarray}
\label{E1}
&& E_{1}\equiv3FH^{2}+3H\dot{F}-\frac{\omega}{2}\dot{\phi}^{2}-V(\phi)=0,\\
\label{E2}
&& E_{2}\equiv3FH^{2}+2H\dot{F}+2F\dot{H}+\ddot{F}+\frac{\omega}{2}\dot{\phi}^{2}-V(\phi)=0,\\
\label{E3}
&&E_{3}\equiv\omega\ddot{\phi} + 3\omega H\dot{\phi}
+\frac{1}{2}\dot{\phi}^{2}\omega'_{\phi}+V'_{\phi}-6H^{2}F'_{\phi}-3\dot{H}F'_{\phi}= 0,
\end{eqnarray}
where a dot represents a derivative with respect to the cosmic time $t$, $H \equiv \dot{a}/a$ is the Hubble parameter and prime denotes a derivative wrt the scalar field, for example: $F'_{\phi} = \partial F/\partial \phi $.

From Bianchi identities one has
\begin{equation}
\label{Bianchi}
\dot{\phi}E_{3}+\dot{E}_{1}+3H(E_{1}-E_{2})=0,
\end{equation}
thus, only two of the equations (\ref{E1})--(\ref{E3}) are independent.
For this reason, the scalar field equation (\ref{E3}) can be derived from the equations (\ref{E1})--(\ref{E2})
these are completely describe the cosmological dynamics and we will deal with the GSTG gravity equations only
\begin{eqnarray}
\label{EFR1}
&&3FH^{2}+3H\dot{F}=\frac{\omega(\phi)}{2}\dot{\phi}^{2}+V(\phi),\\
\label{EFR2}
&&H\dot{F}-2F\dot{H}-\ddot{F}=\omega(\phi)\dot{\phi}^{2}.
\end{eqnarray}

We will refer to equations (\ref{EFR1})--(\ref{EFR2}) as for {\it GSTG cosmology} equations.

If $F=1$ equations (\ref{EFR1})--(\ref{EFR2}) are reduced to those for scalar field Friedmann (inflationary) cosmology from GR
\begin{equation}
\label{ER1}
3H^{2}=\frac{\omega(\phi)}{2}\dot{\phi}^{2}+V(\phi),
\end{equation}
\begin{equation}
\label{ER2}
\omega(\phi)\dot{\phi}^{2}=-2\dot{H},
\end{equation}
%%%
where we can chose $\omega=1$ or redefine a scalar field as $\psi=\int\sqrt{\omega(\phi)}d\phi$.

Therefore, one can consider {\it GR cosmology} equations (\ref{ER1})-(\ref{ER2}) as a partial case of GSTG dynamic equations (\ref{EFR1})-(\ref{EFR2}). The partial solutions of the equations (\ref{ER1})-(\ref{ER2}) are de Sitter ones, namely $\phi=const$, $H=\lambda=const$, $V=3\lambda^{2}=const$ which correspond to exponential expansion of the universe.
In our approach, we associate these solutions with Einstein gravity and consider the non-minimal coupling of a scalar field and curvature as the source of the deviations from de Sitter cosmological model.

\section{The quadratic connection between Hubble parameter and coupling function }\label{sect3}

The quadratic connection between Hubble parameter $H(\phi)$ and the function $F(\phi)$ of non-minimal interaction to gravity fist time have been proposed in the form $H=\lambda\sqrt{F}$ with $\lambda > 0$ \cite{Fomin:2017sbt} . Such approach gave possibility for detailed study of the inflation with power-law Hubble parameter.

 Let us consider the generalization of such method with representation of the GSTG cosmology equations based on the following connection: $H=\pm \lambda\sqrt{F}$ with $H(t)=f(t)+\lambda$, where $\lambda>0$ for positive sign and $\lambda<0$ for negative sign; $f(t)$ is $ \mathcal{C}^2$ function of the cosmic time.
As the result, one can derive from the definition of $F(t) $ and equations \eqref{EFR1}-\eqref{EFR2} the following system
\begin{equation}
\label{DER1}
F(t)=\left(1+\frac{f(t)}{\lambda}\right)^2,
\end{equation}
\begin{equation}
\label{DER2}
V(\phi(t))=\frac{1}{\lambda^{2}}\left[3(f(t)+\lambda)^{4}+6(f(t)+\lambda)^{2}\dot{f}
+\dot{f}^{2}+(f(t)+\lambda)\ddot{f}\right],
\end{equation}
\begin{equation}
\label{DER3}
\omega(\phi(t))\dot{\phi}^{2}=-\frac{2}{\lambda^{2}}\left[\dot{f}^{2}+(f(t)+\lambda)\ddot{f}\right].
\end{equation}

As a very wide  inflationary evolution of the scale factor we consider the exponential-power law (EPL) expansion $H(t)=\frac{m}{t}+\lambda$ or $f(t)=\frac{m}{t}$. Let us describe the parameters of cosmological models for this case.

\subsection{The parameters of cosmological models for EPL expansion}
We may find the solutions in terms of the cosmic time for the EPL expansion from equations \eqref{DER1}-\eqref{DER3}
\begin{equation}
\label{SOL1}
F(\phi(t))=\sum^{2}_{k=0}A_{k}t^{-k}=F_{0}+F_{1}+F_{2},
\end{equation}
where $A_{0}=1$, $A_{1}=\frac{2m}{\lambda}$, $A_{2}=\frac{m^{2}}{\lambda^{2}}$.
\begin{equation}
\label{SOL2}
X=-\frac{1}{2}\omega(\phi(t))\dot{\phi}^{2}=\sum^{4}_{k=3}B_{k}t^{-k}=X_{3}+X_{4},
\end{equation}
where $B_{3}=\frac{2m}{\lambda}$, $B_{4}=\frac{3m^{2}}{\lambda^{2}}$.
\begin{equation}
\label{SOL3}
V(\phi(t))=\sum^{4}_{k=0}C_{k}t^{-k}=V_{0}+V_{1}+V_{2}+V_{3}+V_{4},
\end{equation}
where
\begin{eqnarray*}
&&C_{0}=3\lambda^{2},~~~~C_{1}=12m\lambda,~~~~C_{2}=6m(3m-1),\\
&&C_{3}=\frac{2m}{\lambda}(6m^{2}-6m+1),~~~C_{4}=\frac{3m^{2}}{\lambda^{2}}(m^{2}-2m+1).
\end{eqnarray*}

%%%%%%-----NEW
Therefore, one can consider the inflation on two time scales: from $k=0$ to $k=2$ which is governed by the non-minimal coupling $F(\phi)\gg V(\phi)$ for $|\lambda|\ll1$ and from $k=3$ to $k=4$ which is governed, in general case, by the kinetic energy $X$ and the potential $V(\phi)$. Also, we note that the condition $m\gg1$ corresponds to the regime of the potential dominance $V(\phi)\gg X$.
This division into different time intervals corresponds to the inflationary stage only in the considered models. From the proposed analysis it follows that the difference between this type of inflation and standard one~\cite{Starobinsky:1980te,Guth:1980zm,Linde:1981mu,Albrecht:1982wi,Chervon:2019sey} is the presence of the stage of predominance of a non-minimum coupling at the beginning of the inflation.
Now, we consider the exponential-power-law (EPL) cosmological dynamics including the stages beyond the initial inflationary one on the basis of the analysis of universe's relative acceleration.
%%%%%%%%%%%%---end NEW

%%%%%%%%%--SV-version
It is interesting to note, that when $k$ runs from $k=0$ to $k=2$ we have non-minimal coupling $F$ in the action \eqref{actionFR}
$S_{(STG)}$ but the kinetic part $ X=-\frac{1}{2}\omega(\phi(t))\dot{\phi}^{2}$ is absent. In the case when $k$ runs from $k=3$ to $k=4$ we have the kinetic part $ X$ but non-minimal coupling $F$ is absent. It is evident that the potential is present for all $k$.

%%%%%%%%%% -- end SV-version

\subsection{Accelerated period in EPL inflation}

The scale factor of EPL inflation is
\begin{equation}\label{a-epl}
  a(t)=a_s t^m e^{\lambda t},~a_s=const.
\end{equation}

Then, the relative acceleration $Q(t)$ is
\begin{equation}\label{aa-epl}
 Q(t):= \frac{\ddot{a}(t)}{a(t)}= m(m-1)z^2+2\lambda m z +\lambda^2,~~z=1/t
\end{equation}
 The roots of the quadratic trinomial of rhs are
 \begin{equation}\label{roots}
z_1=-\frac{\lambda}{\sqrt{m}+m},~~  z_2=\frac{\lambda}{\sqrt{m}-m},
\end{equation}
That is for cosmic time $t$ we have
\begin{equation}\label{t-roots}
t_1=-\frac{\sqrt{m}+m}{\lambda},~~  t_2=\frac{\sqrt{m}-m}{\lambda}.
\end{equation}

To analyze the dynamic, taking into account the condition $|\lambda|\ll1$,
we will consider the normalized constant parameter $\tilde{\lambda}=\lambda\times 10^{-6}$.

Positive acceleration will be in the following cases:
 \begin{itemize}
 \item $\tilde{\lambda}>0,~m>1$, then $ z_1<0,~z_2<0$, and for positive $z$ or $t$ acceleration period is $ z\in (0, \infty)$. Thus we have eternal inflation during the period $t \in (0, \infty)$.
 \item   $ \tilde{\lambda}>0,~0<m<1$ then  $ z_1<0,~z_2>0$.  We have an acceleration  during the period $ z \in (0, z_2)$. It means that we have acceleration during the period $ t \in (t_2, \infty)$. % where $t_2=\sqrt{m}(1-\sqrt{m})/\lambda$.
 \item $\tilde{\lambda}<0,~m>1$. We have two periods of accelerated expansion:
 $ z\in (0, z_1)$ which corresponds to $ t \in (t_1, \infty)$ and
 $ z\in (z_2, \infty)$ or $t \in (0, t_2)$. %Here $t_1 = -\sqrt{m}(1+\sqrt{m})/\lambda$.
 \item $ \tilde{\lambda}<0,~0<m<1$. We have $z_1>0,~z_2<0$ and $t_1>0,~t_2<0$ correspondingly. For positive $t$ we have accelerating period $ z\in (0, z_1)$ which corresponds to $ t \in (t_1, \infty)$.
 \end{itemize}

\begin{figure}[ht]%\label{a-epl-1}
\begin{center}
{\includegraphics*[scale=0.5]{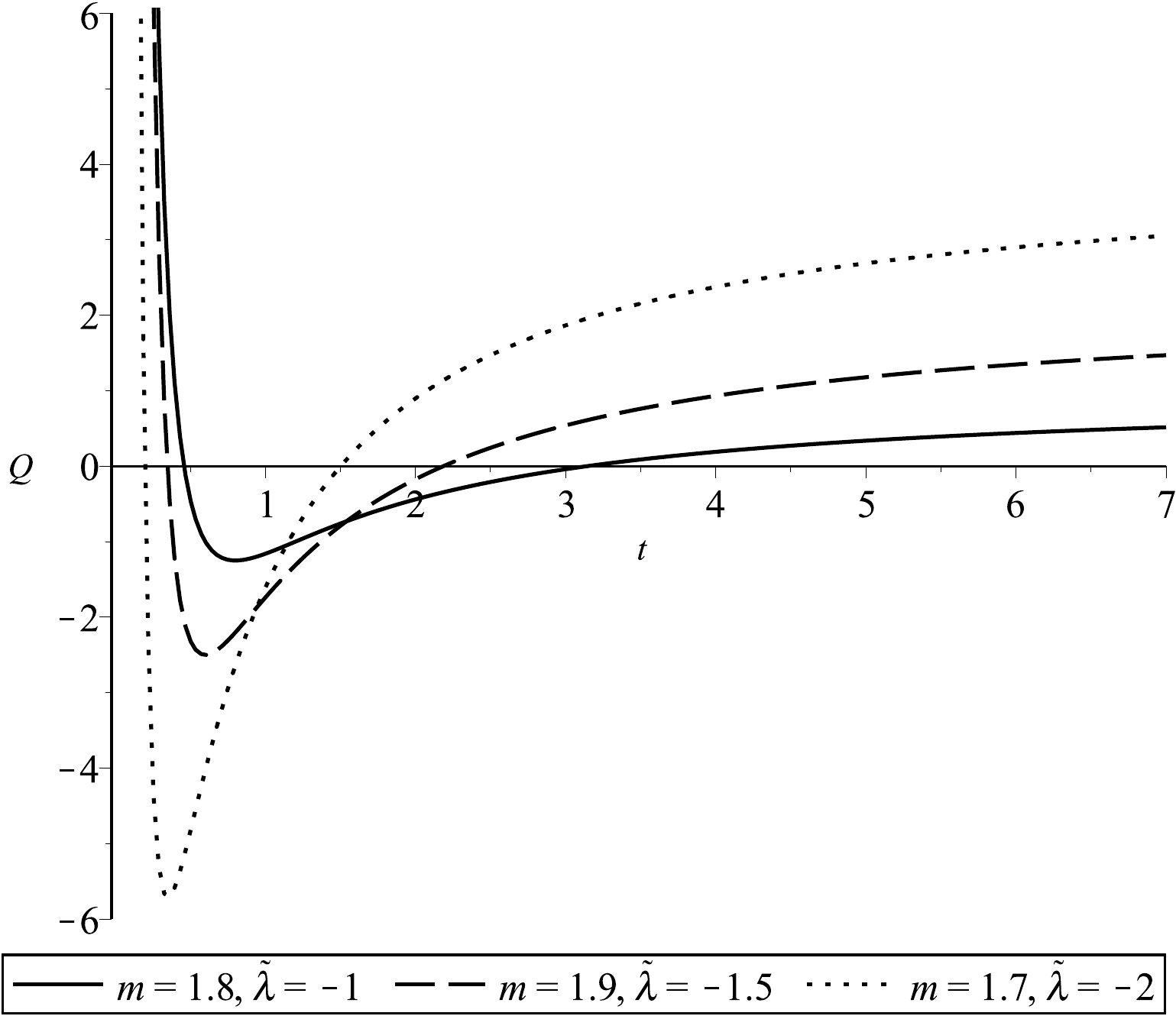}}
\end{center}
\caption{The relative acceleration $Q$ of the universe's expansion for different values of the parameters $\tilde{\lambda}=\lambda\times 10^{-6}$ and $m$.}
\label{a-epl-1}
\end{figure}

It is interesting to note the case when we have two accelerating periods, i.e. when $\tilde{\lambda}<0,~m>1$. Effectively the early inflation takes place for $\tilde{\lambda}=-1,~m=1.8$ during the period $t\in (0,~0.458)$; the later inflation  -- when $t\geq 3.142.$
%%%%

On FIG. \eqref{a-epl-1} one can see two stages of universe acceleration (for different parameters $m$ and $\tilde{\lambda}$), and that the expansion of the early inflation is faster than the latter one. Between these two stages, the universe expands without acceleration.
Thus, the EPL-dynamics (\ref{a-epl}) with $\tilde{\lambda}<0$ and $m>1$ corresponds to the correct change in the stages of the universe's expansion.
%%%%
%-----------------

\begin{figure}[ht]
\begin{center}
\includegraphics*[scale=0.5]{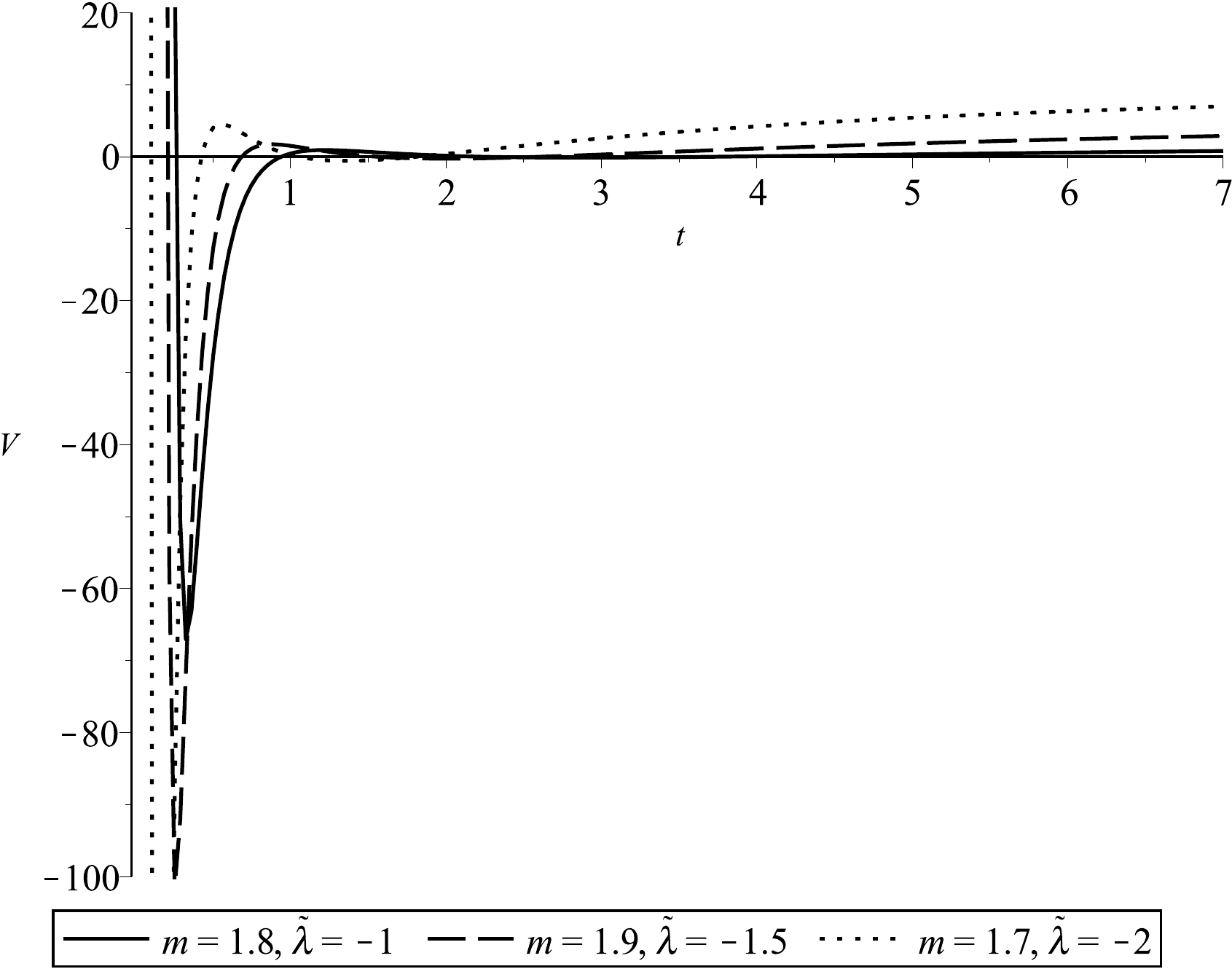}
\end{center}
\caption{Potential energy for different values of $\tilde{\lambda}<0$ and $m>1$.}
\label{epl_V_t}
\end{figure}

\begin{figure}[ht]
\begin{center}
\includegraphics*[scale=0.5]{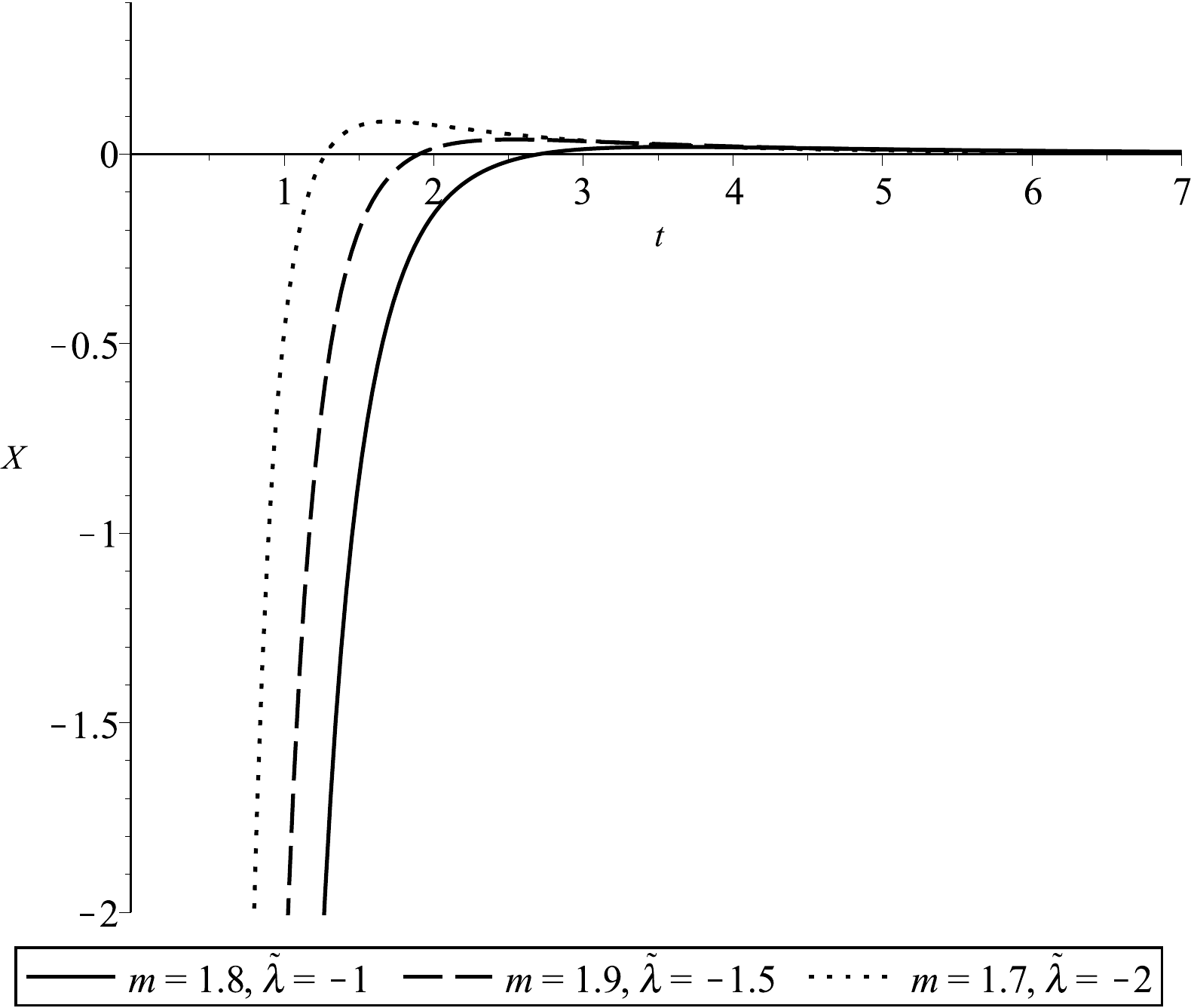}
\end{center}
\caption{Kinetic energy for different values of $\tilde{\lambda}<0$ and $m>1$.}
\label{epl_X_t}
\end{figure}

\begin{figure}[ht]
\begin{center}
{\includegraphics*[scale=0.5]{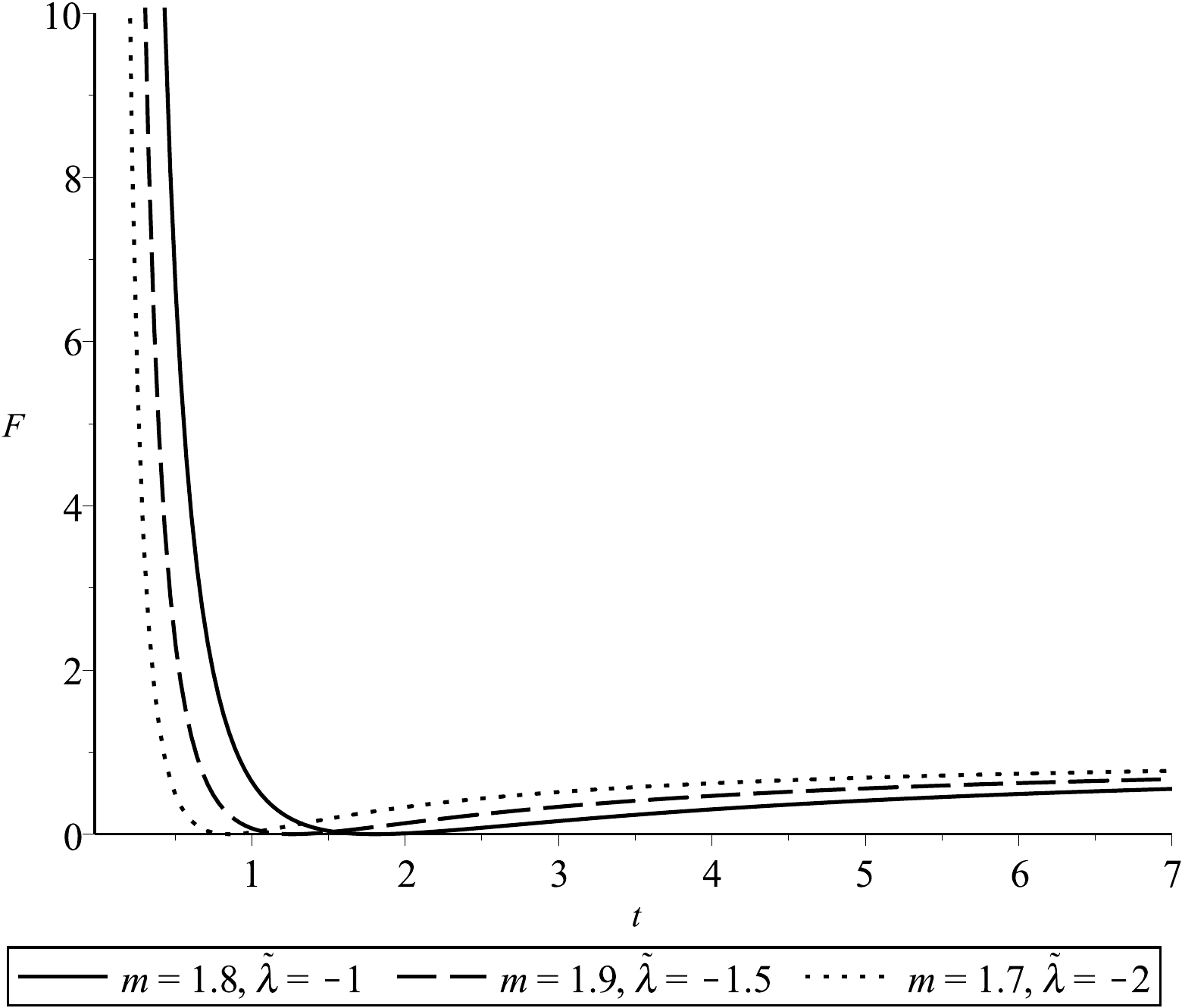}}
\end{center}
\caption{$F(t)$ for different values of $\tilde{\lambda}<0$ and $m>1$.}
\label{epl_F_t}
\end{figure}

The dependencies of potential and kinetic energies on time are described by relations \eqref{SOL2} and \eqref{SOL3}.
Graphical representations  on FIG. \eqref{epl_V_t} and \eqref{epl_X_t} show the behavior of potential and kinetic energies for different values of $\tilde{\lambda}$ and $m$ during both inflationary periods. One can see that during the deceleration period the kinetic energy is crossing the phantom zone. FIG. \eqref{epl_F_t} demonstrates influence of non-minimal interaction. %during early stage of universe evolution.
Let us note that for early inflation ($t\in (0,~0.458)$, for $\tilde{\lambda}=-1,~m=1.8$) the influence of non-minimal interaction is very high. For the later inflation the $F(t)$ becomes much less.

\section{The exact solutions for generalized polynomial potential}\label{sect4}

Now, we consider the following equation
\begin{equation}
\label{Function}
t^{-1}\equiv\Psi(\phi),
\end{equation}
where $\Psi(\phi)$ is the some function of a scalar field.
The evolution of a scalar field $\phi=\phi(t)$ can be defined by the choice of function $\Psi(\phi)$.

Thus, form the equation (\ref{SOL3}) one has the generalized polynomial potential
\begin{equation}
\label{SOL3G}
V(\phi)=3\lambda^{2}+\delta V(\phi),~~~~~\delta V(\phi)\equiv\sum^{4}_{k=1}C_{k}[\Psi(\phi)]^{k},
\end{equation}
with corresponding coupling function which we define from (\ref{SOL1})
\begin{equation}
\label{SOL1G}
F(\phi)=1+\delta F(\phi),~~~~~\delta F(\phi)\equiv\sum^{2}_{k=1}A_{k}[\Psi(\phi)]^{k},
\end{equation}
and kinetic function
\begin{equation}
\label{SOL2G}
\omega(\phi)=-\frac{2}{U(\phi)}\sum^{4}_{k=3}B_{k}[\Psi(\phi)]^{k},~~~~~\dot{\phi}^{2}\equiv U(\phi),
\end{equation}
which can be obtained from the equation  (\ref{SOL2}).

The term $\delta V$ in potential (\ref{SOL3G}) can be interpreted as corrections in the low-energy limit of the effective theory of quantum gravity~\cite{Baumann:2014nda,Martin:2013tda} which induced in this case by the non-minimal coupling $\delta F(\phi)$ of a scalar field and curvature.

Also, we note that the potential of the scalar field depends on the rate of the expansion of the universe, namely
\begin{itemize}
\item for $m=0$ one has the flat potential $V=V_{0}=3\lambda^{2}=const$,
\item for $m=1/3$ one has $V_{2}=0$,
\item for $m=\frac{1}{2}\left(1\pm\frac{1}{\sqrt{3}}\right)$ one has $V_{3}=0$,
\item for $m=1$ one has $V_{4}=0$,
\item for other values of the parameter $m$ all components $V_{k}\neq0$.
\end{itemize}

Thus, one can generate the exact cosmological solutions for the case of the generalized polynomial potential by the choice of the generating function $\Psi(\phi)$. Let's consider some exact solutions for the inflationary models based on this approach.

\subsection{The model with polynomial potential}

For this case, we consider the following generating function
\begin{equation}
\label{GenFunc}
\Psi(\phi)=\left(\frac{\phi}{\alpha}\right)^{1/\beta},
\end{equation}
where $\alpha$ and $\beta$ are arbitrary constants.

From equation (\ref{Function}) one can find the evolution of the scalar field and the function $U(\phi)=\dot{\phi}^{2}$ form it's definition
\begin{equation}
\phi(t)=\alpha t^{-\beta},~~~~~U(\phi)=-\alpha\beta\left(\frac{\phi}{\alpha}\right)^{\frac{1+\beta}{\beta}}.
\end{equation}

From the expression for generating function (\ref{GenFunc}) and equation (\ref{SOL3G}) one has the polynomial potential
\begin{equation}
\label{POTQC}
V(\phi)=3\lambda^{2}+\sum^{4}_{k=1}C_{k}\left(\frac{\phi}{\alpha}\right)^{k/\beta},
\end{equation}
non-minimal coupling function from the equation (\ref{SOL1G})
\begin{equation}
\label{COUPQC}
F(\phi)=\left[1+\frac{m}{\lambda}\left(\frac{\phi}{\alpha}\right)^{1/\beta}\,\right]^{2}=
1+\sum^{2}_{k=1}A_{k}\left(\frac{\phi}{\alpha}\right)^{k/\beta},
\end{equation}
and the kinetic function from (\ref{SOL2G})
\begin{equation}
\label{KINQC}
\omega(\phi)=\frac{2}{\alpha\beta}\sum^{4}_{k=3}B_{k}\left(\frac{\phi}{\alpha}\right)^{\frac{k-\beta-1}{\beta}}.
\end{equation}

Polynomial potentials (\ref{POTQC}) are considered in the context of the effective quantum gravity theory in the works ~\cite{Baumann:2014nda,Martin:2013tda}.
Thus, we reconstructed the parameters (\ref{COUPQC})--(\ref{KINQC}) of the scalar-tensor gravity theory for this type of potentials in the inflationary models under consideration.

\subsection{The model with exponential potential}

Let us study logarithmic evolution of the scalar field which is often appeared in standard Friedmann cosmology \cite{Chervon:2017kgn}
\begin{equation}\label{log-t}
\phi(t)=\beta \ln t,
\end{equation}
this case corresponds to the choice of the following generating function $\Psi(\phi)=\exp\left(-\phi/\beta\right)$ and
$U(\phi)=\beta^{2}\exp\left(-2\phi/\beta\right)$.

With suggestion \eqref{log-t} we have the expansion on exponents with the same coefficients as in equations \eqref{SOL1}-\eqref{SOL3}:
\begin{equation}
\label{EXPPOT}
V(\phi)=3\lambda^{2}+\sum^{4}_{k=1}C_{k}\exp \left(-\frac{k}{\beta}\phi \right),
\end{equation}
non-minimal coupling function
\begin{equation}
F(\phi)=1+\sum^{2}_{k=1}A_{k}\exp \left(-\frac{k}{\beta}\phi \right),
\end{equation}
and kinetic function
\begin{equation}
\omega(\phi)=-\frac{2}{\beta^{2}}\sum^{4}_{k=3}B_{k}\exp \left(\frac{2-k}{\beta}\phi \right).
\end{equation}

The model of quintessential inflation with the potential (\ref{EXPPOT}) on the basis of Einstein gravity was considered earlier in the paper~\cite{Barreiro:1999zs}.

\section{Inflationary stage for physically motivated potential}\label{sect6}

Now we consider another view on the problem, different from that described in section \ref{sect4} where we were started, in fact, from known evolution of a scalar field. Here we changed the situation and looking for algorithm which give us possibility to find parameters of scalar-tensor gravity which give exact solutions for given physical potential.  The algorithm is demonstrated on the Higgs potential in the following subsection. Note that the form of the potential does not influence on the function of non-minimal coupling $F(t)$.

Due to the fact that the scalar field potential is of key importance for determining physical processes at the stage of cosmological inflation, it is precisely the potential $V(\phi)$ that is specified for constructing models of the early universe. The form of the scalar field potential is determined from elementary particle physics, quantum field theory, theories of unifying fundamental interactions, such as supersymmetric theories and string theories in the context of the inflationary paradigm~\cite{Baumann:2014nda}. We will call the potentials of a scalar field associated with these mechanisms as {\it physical potentials}. Physical mechanisms corresponding to a large number of inflationary potentials were considered in the review~\cite{Martin:2013tda}.

\subsubsection{The solution for Higgs potential}

Let us use Eq. \eqref{SOL3} to find the scalar field dependence on time. The Higgs potential we take in the standard physical form~\cite{Martin:2013tda,Mishra:2018dtg}:
\begin{equation}\label{higgsV}
 V(\phi)=\frac{\lambda_H}{4}\left( \phi^2-v^2 \right)^2,
\end{equation}
where $\lambda_{H}$ is the Higgs coupling constant and $v$ is the vacuum expectation value of the Higgs field. 

Denoting right hand side of  \eqref{SOL3} as $T(1/t)$ one can obtain
\begin{equation}\label{phi-higgs}
 \phi= \pm \sqrt{v^2+\frac{2}{\sqrt{\lambda_H}}\sqrt{T(1/t)}}
\end{equation}

\begin{figure}[ht]
\begin{center}
{\includegraphics*[scale=0.5]{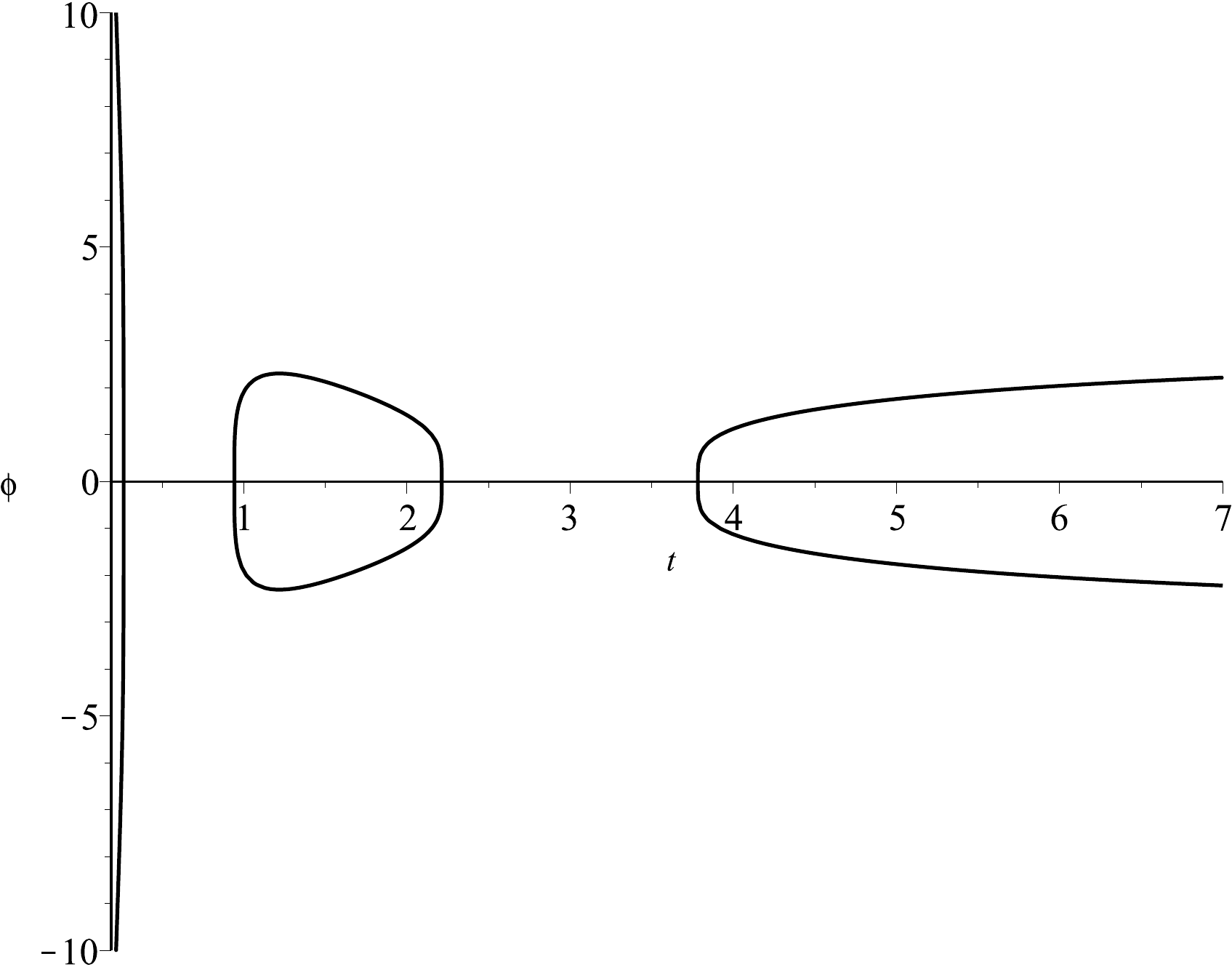}}
\end{center}
\caption{Evolution of scalar field for Higgs potential for  $\tilde{\lambda}=-1$, $m=1.8$. Zeros of the $T(1/t)$ are: 0.263, 0.941, 2.213, 3.783.}
\label{higgs_phi_t}
\end{figure}

This solution for the scalar field is demonstrated on the FIG. \eqref{higgs_phi_t}. Because of the plus-minus sign before the square root we have two branches in the picture and gaps, when the function under square root is negative.

Then, calculating $\dot{\phi}^2$
$$
\dot{\phi}^2=\frac{1}{4\lambda_H T(1/t)}\left(v^2+ \frac{2}{\sqrt{\lambda_H}}\sqrt{T(1/t)}\right)^{-1} \dot{T}^2
$$
and inserting the result in \eqref{SOL2} one can define the kinetic function $\omega(\phi(t))$.

\begin{figure}[ht]
\begin{center}
{\includegraphics*[scale=0.5]{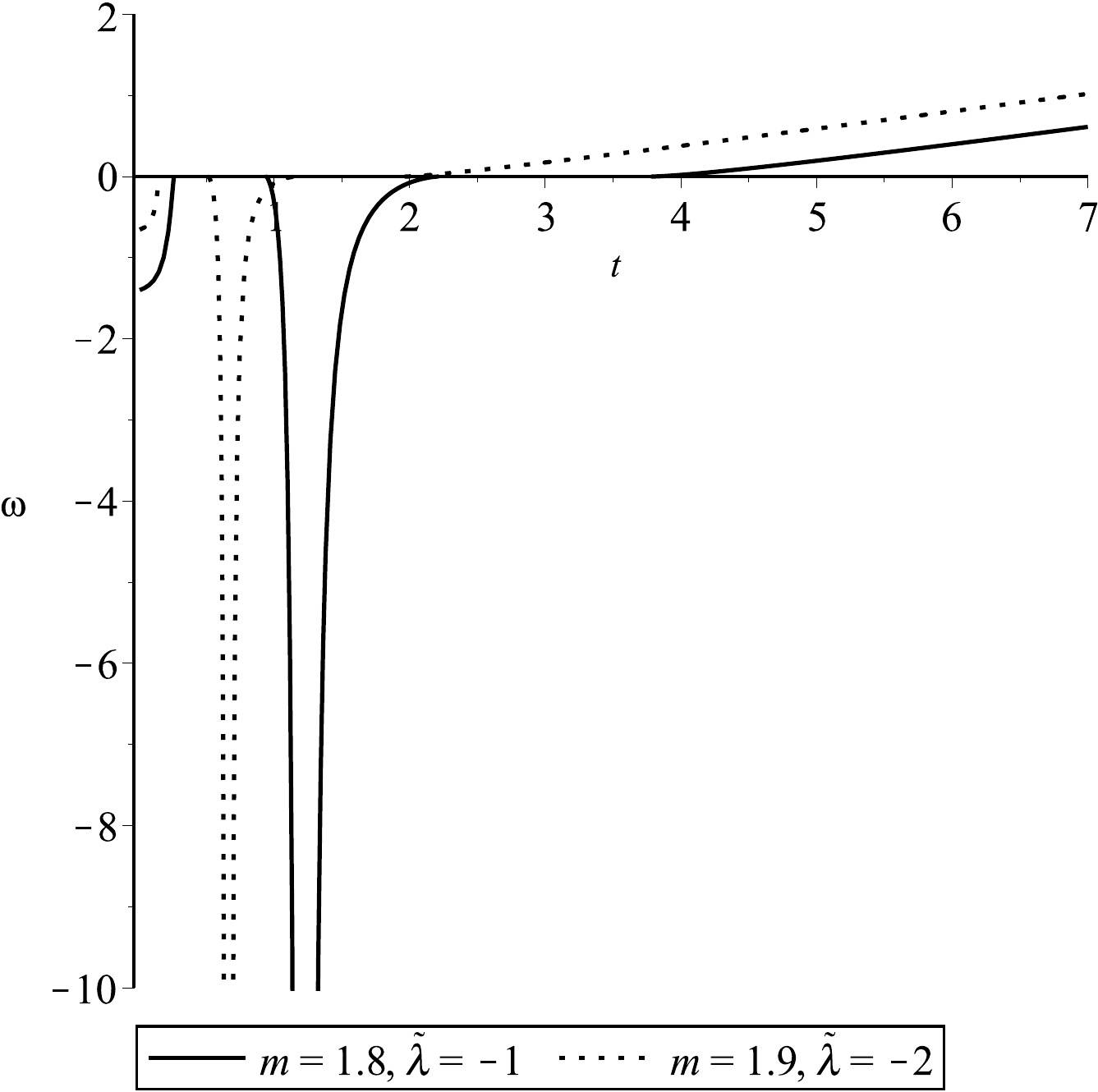}}
\end{center}
\caption{Kinetic function $\omega$ for different values of $\tilde{\lambda}<0$ and $m>1$.}
\label{higgs_omega_t}
\end{figure}

FIG.\eqref{higgs_omega_t} demonstrates the change of kinetic function $\omega $ during allowable periods for real solutions of the scalar field and for different values of $\tilde{\lambda}<0$ and $m>1$.
The function $\omega (t)$ (when $m=1.8$, $\tilde{\lambda} = -1$) tends to $ - \infty $ when $t \rightarrow 1.216 $.

\subsubsection{The solution for Higgs-Starobinsky potential}
We consider approximated Higgs potential \cite{Bezrukov:2007ep,Bezrukov:2008ej,Martin:2013tda,Mishra:2018dtg}
%(see, for ex., [topor-18])
\begin{equation}\label{star}
V(\phi)=V_0\left(1-\exp \left[-\frac{2\phi}{\sqrt{6}M_P}\right] \right)^2,~~~V_0=const,
\end{equation}
which is the analog of Starobinsky one (see,  \cite{Mishra:2018dtg} and literature quoted therein)
\begin{equation}\label{V-Star}
 V(\phi)=\frac{3}{4}m^2_{\ast}M_P^2\left(1-\exp \left[-\frac{2\phi}{\sqrt{6}M_P}\right] \right)^2,
\end{equation}
where $m_{\ast}$ is the mass of a scalar field, also, we consider $M^{-2}_P=1$ in chosen system of units.

Once again, equating the potential \eqref{star} to $T(1/t)$ we have
\begin{equation}\label{phi-st}
\phi=-\sqrt{3/2}\ln\mid 1 \mp \sqrt{T(1/t)/V_0}\mid
\end{equation}

\begin{figure}[ht]
\begin{center}
{\includegraphics*[scale=0.5]{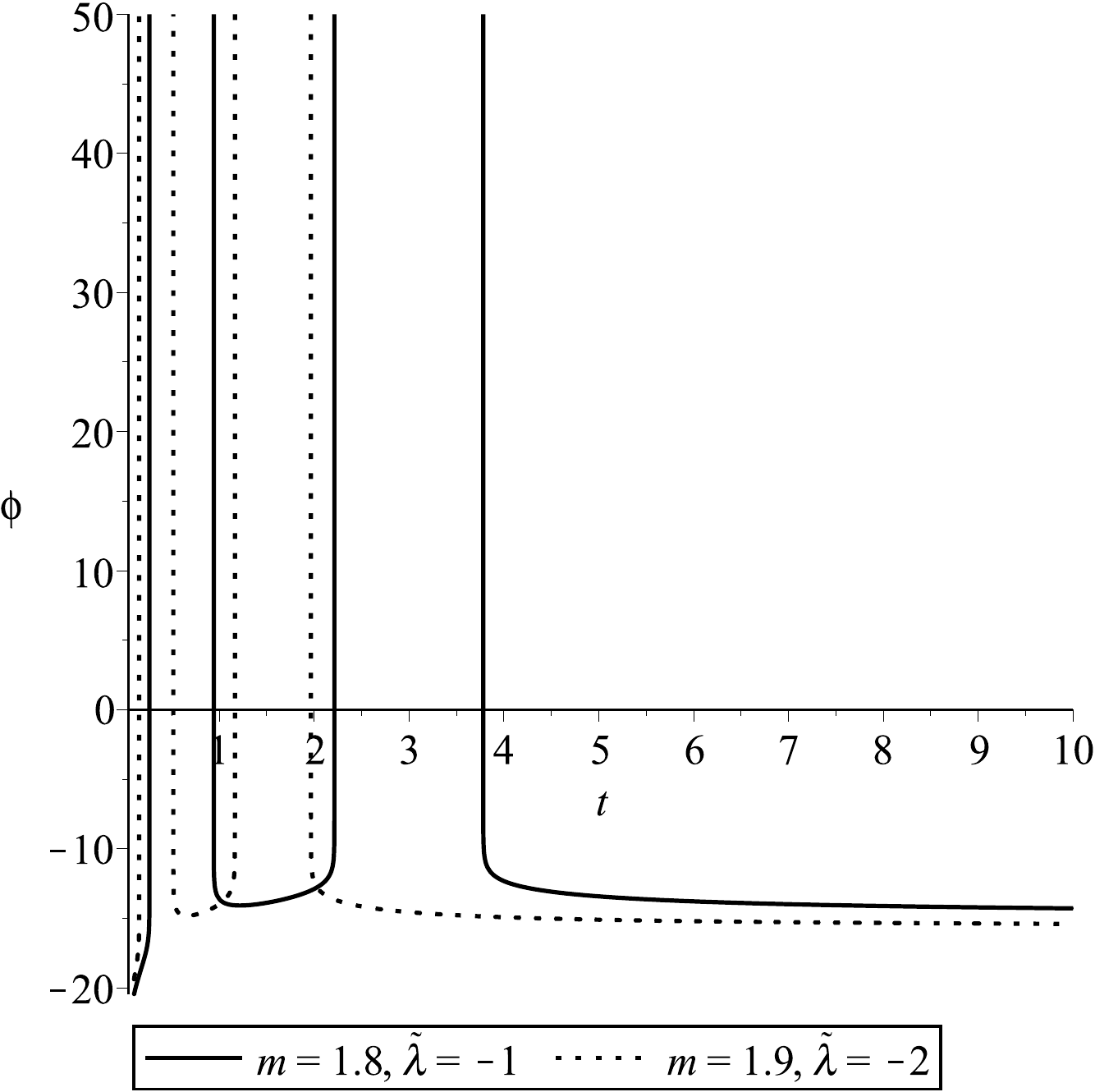}}
\end{center}
\caption{Evolution of scalar field for Higgs-Starobinsky potential for different values of $\tilde{\lambda}<0$ and $m>1$.}
\label{star_phi_t}
\end{figure}

The general form of Higgs-Starobinsky potential is well-known and we displayed on FIG.\eqref{star_phi_t} the evolution of scalar field for different values of $\tilde{\lambda}<0$ and $m>1$. Note that during later evolution the scalar field %is almost constant.
varies very slowly.

Because there are negative zone for square root in the logarithm's argument  we get few gaps and function is tending to infinity when the logarithm's argument tends to zero.

\subsubsection{The solution for Coleman-Weinberg potential}

The scalar field dependence on $t$ for Coleman-Weinberg potential~\cite{Martin:2013tda}
\begin{equation}
V(\phi)= \alpha \phi^4\left( \ln \left(\frac{\phi}{v_\phi}\right)-\frac{1}{4}\right)+
\frac{\alpha}{4}v_\phi^4
\end{equation}

can be extracted from the equation
\begin{equation}\label{phi-CW}
\phi^4 (\ln \left( \frac{\phi}{v_\phi}\right)-1/4)=T(1/t)/\alpha-
\frac{1}{4}v_\phi^4
\end{equation}

Note, that it is rather difficult to represent the diagram of Coleman-Weinberg potential in physical units.

\subsubsection{Massive scalar field}

The exact solution for massive scalar field for GR cosmology can be represented by the following formulae~\cite{Martin:2013tda}:
\begin{equation}\label{V-msf}
  V(\phi)=\frac{1}{2}m^2\phi^2-V_*
\end{equation}
Equating the potential \eqref{V-msf} to $T(1/t)$ one can find
\begin{equation}\label{phi-msf}
\phi (t) = \pm \frac{\sqrt{2}}{m}\sqrt{T(1/t)+V_*},
\end{equation}

Using our method we can find the generalised scalar-tensor gravity admitting the exact solutions with $V_*=0$, what is impossible for GR cosmology.

\begin{figure}[ht]
\begin{center}
{\includegraphics*[scale=0.5]{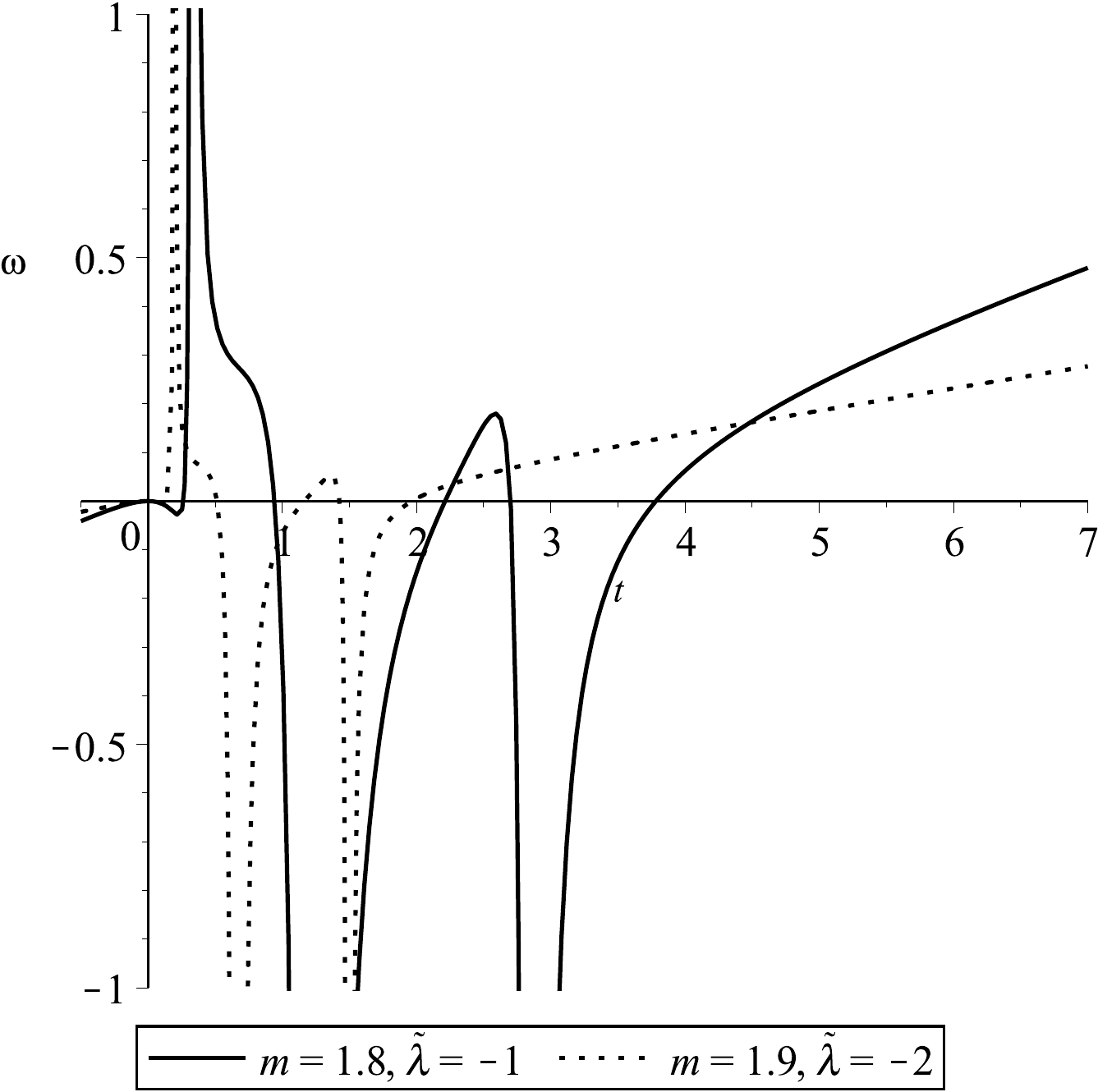}}
\end{center}
\caption{Evolution of massive scalar field for different values of $\tilde{\lambda}<0$ and $m>1$.}
\label{msf_omega_t}
\end{figure}

The function $F(\phi)$ of non-minimal interaction to gravity can not be represented by analytical way. The same relates to kinetic energy $X$ and the kinetic function $\omega $. Evolution of massive scalar field for different values of $\tilde{\lambda}<0$ and $m>1$ is displayed on FIG.\eqref{msf_omega_t}.

\section{Parameters of cosmological perturbations}\label{sect5}

Now, we calculate the parameters of cosmological perturbations to check the correspondence of the inflationary models under consideration to the observational constraints which obtained from the PLANCK observations~\cite{Ade:2015xua} of the cosmic microwave background (CMB) radiation.

In accordance with the theory of cosmological perturbations, during the stage of inflation, quantum fluctuations of the scalar field generate the corresponding perturbations of the space-time metric. In the linear order of perturbation theory, the observed anisotropy and polarization of the CMB are explained by the action of two types of perturbations, namely, scalar and tensor perturbations (relic gravitational waves). The third type of perturbations, namely vector perturbations quickly decay in the process of accelerated expansion of the early universe.

Thus, an estimation of the anisotropy of CMB gives corresponding restrictions on the values of the spectral parameters of cosmological perturbations, which according to modern observations are~\cite{Ade:2015xua}
\begin{align}
\label{PLANCKA1}
&{\mathcal P}_{S}=2.1\times10^{-9},\\
\label{PLANCKA1rn}
&n_{S}=0.9663\pm 0.0041,\\
\label{PLANCKA2rn2}
&r<0.065~~~\text{(Planck 2018/BICEP2/Keck-Array)}.
\end{align}

In the paper~\cite{Fomin:2017sbt} the expressions of the parameters of cosmological perturbations for the inflationary models under consideration on the crossing of Hubble radius ($k=aH$) were obtained in the following form
\begin{equation}
\label{flowPS}
{\mathcal P}_{S}=\frac{\lambda^{2}}{16\pi^{2}\epsilon(\epsilon-\delta)},~~~~~{\cal P}_{ T}=\frac{2\lambda^{2}}{\pi^{2}},~~~
r\simeq32\epsilon(\epsilon-\delta),
\end{equation}
\begin{equation}
\label{flownS}
n_{S}\simeq1+2\epsilon\delta-4\epsilon+2\delta+(1-\epsilon)\left(\frac{\epsilon\delta-\xi}{\epsilon-\delta}\right),~~~~~n_{T}=0,
\end{equation}
where
\begin{equation}
\label{SLP1}
\epsilon=-\frac{\dot{H}}{H^{2}},~~~
\delta=-\frac{\ddot{H}}{2H\dot{H}},~~~
\xi=\epsilon\delta-\frac{1}{H}\dot{\delta},
\end{equation}
are the slow-roll parameters.

The slow-roll conditions for this type of inflation is satisfied when~\cite{Fomin:2017sbt}
\begin{equation}
\label{slow-roll}
\delta_{F}=\frac{\dot{F}}{HF}=\frac{2\dot{H}}{H^{2}}=-2\epsilon\ll1.
\end{equation}

For EPL inflation $H(t)=\frac{m}{t}+\lambda$ one has
\begin{equation}
\label{slow-roll-par}
\delta=\frac{1}{\left(m+\lambda t\right)},~~~\epsilon=m\delta^{2},~~~\xi=\delta^{2},~~~m>0.
\end{equation}

Thus, we obtain
\begin{equation}
\label{flowPS1}
{\mathcal P}_{S}=\frac{2\lambda^{2}}{\pi^{2}r},~~~~~{\cal P}_{ T}=\frac{2\lambda^{2}}{\pi^{2}},
\end{equation}
\begin{equation}
\label{flowr1}
r\simeq32m\delta^{3}(m\delta-1), ~~~~~n_{S}\simeq1+3\delta-5m\delta^{2},
\end{equation}

From (\ref{flowr1}) we obtain the following dependence
\begin{equation}
\label{rn}
r=\frac{2}{625m^{2}}\left[\left(-3+\sqrt{20m(1-n_{S})+9}\right)^{3}\left(7+\sqrt{20m(1-n_{S})+9}\right)\right].
\end{equation}

\begin{figure}[ht!]
\begin{minipage}[h]{0.43\linewidth}
\center{\includegraphics[width=1\linewidth]{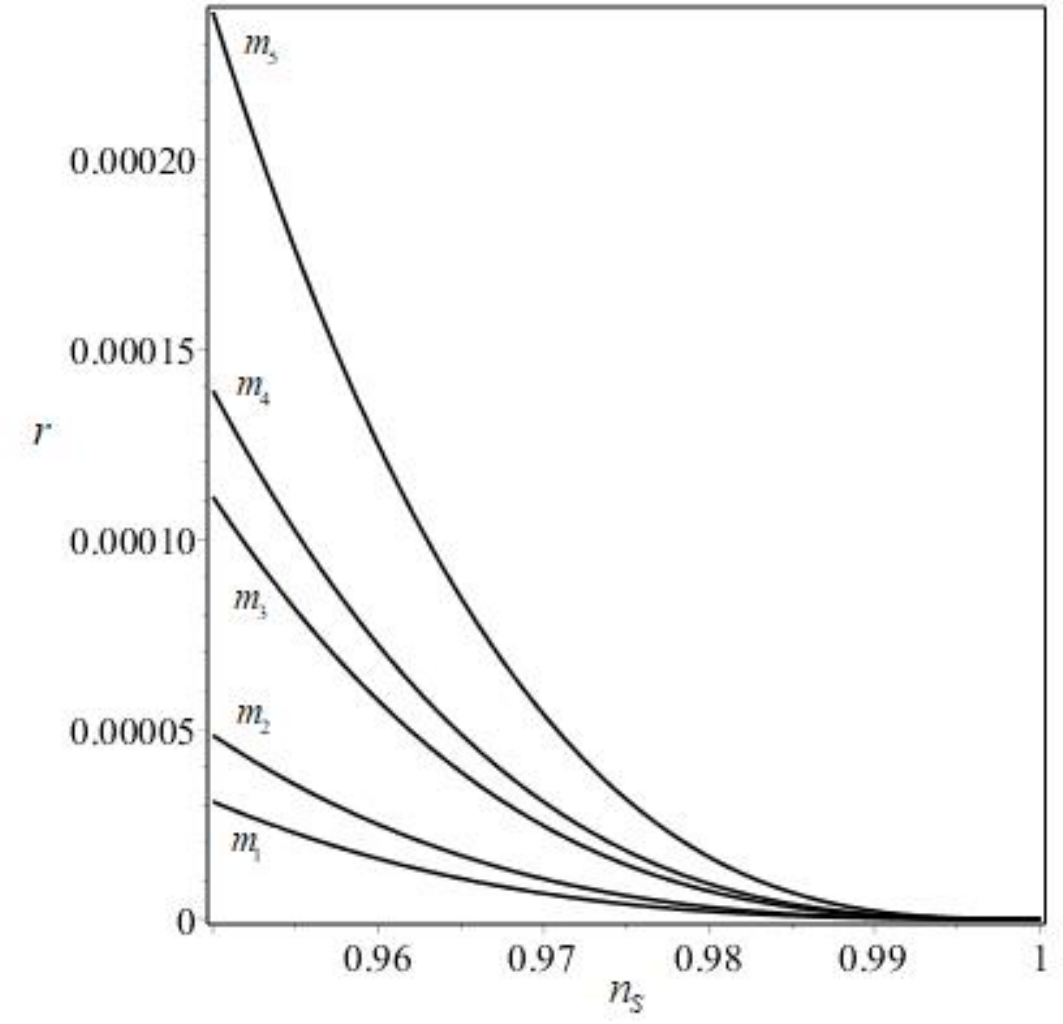}} Different values of $m$\\
\end{minipage}
\hfill
\begin{minipage}[h]{0.43\linewidth}
\center{\includegraphics[width=1\linewidth]{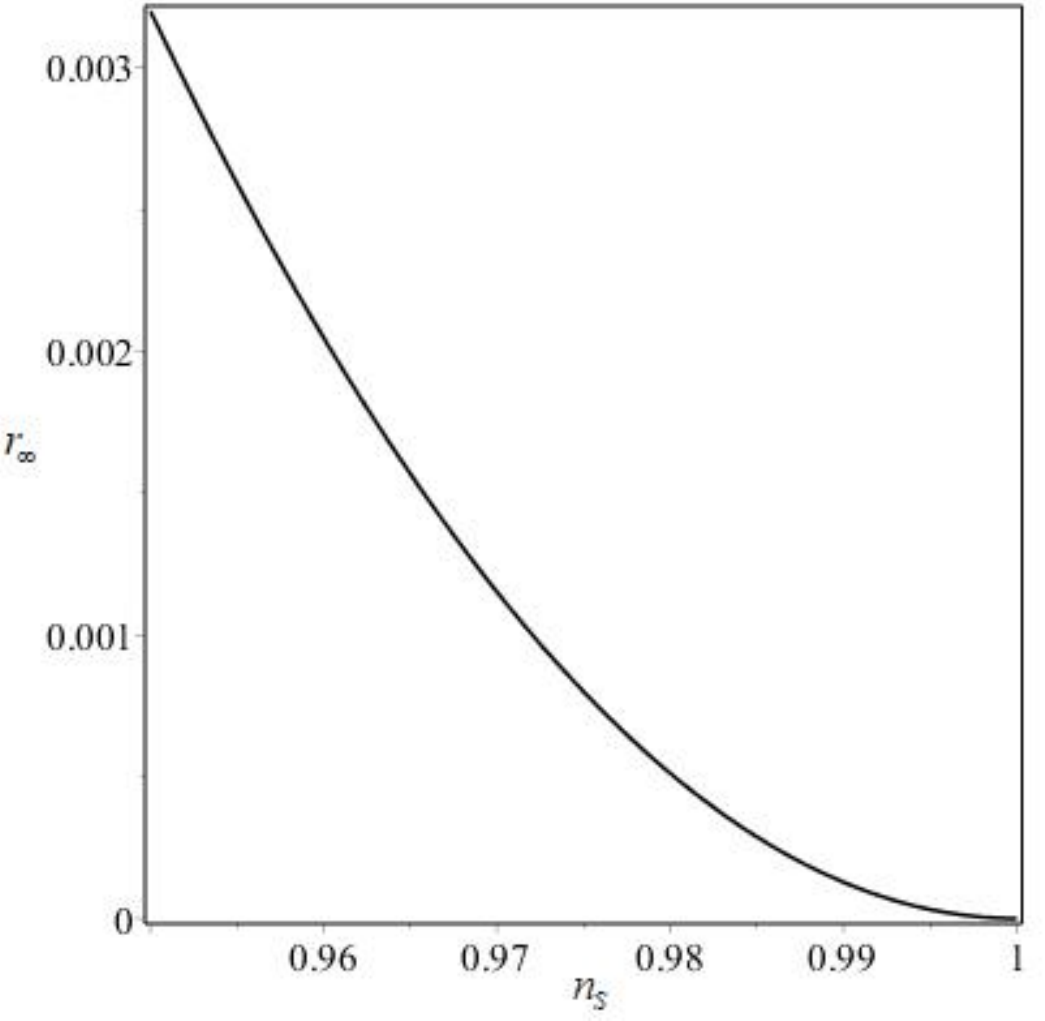}} $m\rightarrow\infty$ \\
\end{minipage}
\caption{The dependence $r=r(n_{S})$ for different values of the constant parameter:
$m_{1}=\frac{1}{2}\left(1-\frac{1}{\sqrt{3}}\right)$,  $m_{2}=\frac{1}{3}$,
$m_{3}=\frac{1}{2}\left(1+\frac{1}{\sqrt{3}}\right)$, $m_{4}=1$, $m_{5}=1.8$ and for $m\rightarrow\infty$.}
\label{fig1}
\end{figure}

Also, from (\ref{rn}) one can obtain that for the spectral index of scalar perturbations $n_{S}=0.9663\pm 0.0041$ the condition $r<0.065$ is satisfied for any positive value of the parameter $m$.
On the Fig.\ref{fig1} the dependence $r=r(n_{S})$ for different values of the constant parameter $m$ is shown.

For $m\rightarrow\infty$ from equation (\ref{rn}) we obtain
\begin{equation}
\label{rnm}
r_{\infty}=\frac{32}{25}\left(n_{S}-1\right)^{2},
\end{equation}
and for the pure de Sitter exponential expansion with $m=0$ tensor-to-scalar ratio $r\rightarrow0$.
Also, we note that from (\ref{slow-roll-par}) one has $\delta(m\rightarrow\infty)=0$ and $\epsilon(m\rightarrow\infty)=0$, i.e.
the slow-roll condition (\ref{slow-roll}) is fulfilled for this case, and, therefore, we can use the relation (\ref{rn}) to calculate tensor-to-scalar ratio.

After substituting the minimal value of the spectral index of scalar perturbations $n_{S}=0.9663- 0.0041=0.9622$ into the relation (\ref{rnm}) one has $r_{\infty(max)}=0.0018$. Therefore, the possible values of the tensor-to-scalar ratio in these models are $r\in(0,0.0018)$ that corresponds to the observational constraint (\ref{PLANCKA2rn2}).

Also, we find the $e$-folds number
\begin{equation}
\label{e-folds}
N=\int^{t_{e}}_{t_{i}}H(t)dt=m\ln\left(\frac{t_{e}}{t_{i}}\right)+\lambda(t_{e}-t_{i})\approx60,
\end{equation}
where $t_{i}$ and $t_{e}$ are the times of the beginning and the end of the inflation.

Thus, tensor-to-scalar ratio in considering models for realistic values of the parameter $m\neq\infty$ can be estimated as $r\sim10^{-4}$ and from the condition ${\mathcal P}_{S}=\frac{2\lambda^{2}}{\pi^{2}r}=2.1\times10^{-9}$ one has $\lambda^{2}\sim10^{-12}$.

Also, we note that the power spectrum of tensor perturbations (relic gravitational waves) is constant on the crossing of Hubble radius
\begin{equation}
\label{powerGW}
{\mathcal P}_{ T}=\frac{2\lambda^{2}}{\pi^{2}}\approx\frac{2}{\pi^{2}}\times10^{-12}=const,
\end{equation}
and amplitude of relic gravitational waves ${\mathcal A}_{ T}={\mathcal P}^{1/2}_{ T}\approx\frac{\sqrt{2}}{\pi}\times 10^{-6}$.

Thus, in this case, the models of cosmological inflation for any potential (\ref{SOL3G}) defined by an arbitrary generating function $\Psi(\phi)$ will satisfy the observational restrictions on the values of cosmological perturbation parameters.

In such a way we showed that the cosmological models based on the quadratic connection between Hubble parameter and coupling function $H=\lambda\sqrt{F}$ with exponential-power-law dynamics $H(t)=\frac{m}{t}+\lambda$ correspond to observational constrains on the parameter of cosmological perturbations for any potential of a scalar field and any expansion rate of the universe.

\section{Conclusion}\label{sect7}

In this paper, we examined cosmological inflationary models with an EPL expansion of the early universe based on scalar-tensor gravity theories implying a non-minimal coupling of the scalar field and curvature. To comply with these models with observational restrictions on the values of the spectral parameters of cosmological perturbations, we considered the quadratic relationship between the non-minimal coupling function and the Hubble parameter $H\propto\sqrt{F}$.

The first result of the proposed approach was the possibility of comparing the values of the potential of the scalar field, its kinetic energy, and the function of non-minimal coupling at different time scales. This feature is based on a polynomial representation of all model's parameters (\ref{SOL1})--(\ref{SOL3}).
This result allows us to determine the nature of the inflationary process by the dominance of some component in the considered models. Namely, we obtained a mode of predominance of non-minimal coupling $F\gg V$ for the times  from $k=0$ to $k=2$ and potential $V\gg X$ for the times from $k=3$ to $k=4$.

The second result is the possibility of constructing exact cosmological solutions for generalized polynomial potentials based on the choice of one generating function $\Psi(\phi)$ from the equations  (\ref{Function})--(\ref{SOL2G}). The examples of exact solutions are given in Sec.\ref{sect4}.
Further, the dependences of the kinetic energy of the scalar field and its potential on the cosmic time for various physical potentials and $m\sim1$ were considered.

The third result is that this class of models corresponds to observational restrictions on the values of the parameters of cosmological perturbations for any potentials of a scalar field and any expansion rate of the universe. Based on observational restrictions on the value of the power spectrum of scalar perturbations and the tensor-to-scalar ratio  (\ref{PLANCKA1})--(\ref{PLANCKA2rn2}), we estimated the value of free model's parameter as $\lambda^{2}\sim10^{-12}$.

Thus, the proposed approach allows one to consider phenomenologically correct models of the early universe and compare the parameters of scalar-tensor gravity theories for different physical potentials, which makes it possible to make a consistent analysis for many inflationary models.

The development of the proposed approach is the analysis of higher orders corrections to the potential
$$
V(\phi)=3\lambda^{2}+\sum^{4}_{k=1}C_{k}[\Psi(\phi)]^{k}+\sum^{\infty}_{k=5}\tilde{C}_{k}[\tilde{\Psi}(\phi)]^{k},
$$
which can be induced by non-minimal coupling of a scalar field with higher curvature terms, for example, with the Gauss-Bonnet scalar~\cite{Kanti:2015pda,vandeBruck:2015gjd,Fomin:2017vae,Fomin:2019yls,Odintsov:2018zhw,Pozdeeva:2019agu}.

Thus, in this approach, the non-minimal coupling of the scalar field and scalar curvature $R$ defines a segment of a more general polynomial potential, which can be reconstructed for cosmological models that satisfy observational constraints on the values of cosmological perturbation parameters and the requirement for the correct change of the stages of universe's expansion.

\begin{acknowledgments}
For this work F.I.V. and S.V.C. are partially supported by the RFBR grant 18-52-45016  IND a.
S.V.C. is grateful for the support of the Program of Competitive Growth of Kazan Federal University.
\end{acknowledgments}

\bibliography{ref}

\end{document}